\documentclass[12pt]{article}
\usepackage{axodraw}
 
\def\ltsim{\lower3pt\hbox{$\, \buildrel < \over \sim \, $}}
\def\gtsim{\lower3pt\hbox{$\, \buildrel > \over \sim \, $}}
\def\be{\begin{equation}}
\def\ee{\end{equation}}
\def\ba{\begin{eqnarray}}
\def\ea{\end{eqnarray}}
\def\ga{\mathrel{\raise.3ex\hbox{$>$\kern-.75em\lower1ex\hbox{$\sim$}}}}
\def\la{\mathrel{\raise.3ex\hbox{$<$\kern-.75em\lower1ex\hbox{$\sim$}}}}
\newcommand{\sect}[1]{\section{#1}\setcounter{equation}{0}}

\begin{document}

\baselineskip=16pt 
\begin{titlepage}
\rightline{OUTP-00-48P}
\rightline{hep-th/0011141}
\rightline{November 2000}  
\begin{center}

\vspace{0.5cm}

\large {\bf A new bigravity model with exclusively positive branes}

\vspace*{5mm}
\normalsize

{\bf Ian I. Kogan\footnote{i.kogan@physics.ox.ac.uk}, Stavros
Mouslopoulos\footnote{s.mouslopoulos@physics.ox.ac.uk} and Antonios Papazoglou\footnote{a.papazoglou@physics.ox.ac.uk}}

\smallskip 
\medskip 
{\it Theoretical Physics, Department of Physics, Oxford University}\\
{\it 1 Keble Road, Oxford, OX1 3NP,  UK}
\smallskip

\vskip0.6in \end{center}

\centerline{\large\bf Abstract}

We propose a new ``bigravity'' model  with two positive tension $AdS_4$ branes in
$AdS_5$ bulk and no negative tension
branes. The bounce of the ``warp'' factor mimics the effect of a
negative brane and thus gives rise to an anomalously light graviton KK
mode. This configuration satisfies  the weak energy condition and has
no ghost state. In
addition,  the extra polarization states  of the
massive graviton practically decouple and thus it does not contradict to
Einsteinian gravity. However, the
model has certain phenomenological  difficulties associated with the
presence of a negative cosmological constant on the branes.

\vspace*{2mm} 

\end{titlepage}

\section{Introduction}

Brane universe models \cite{80s} in more than four dimensions have been extensively studied over the
last three years because they provide novel ideas for resolutions of
long standing problems to particle physics such as the Planck
hierarchy \cite{large,RS1} one. Moreover, mechanisms to
localize gravity on a brane \cite{gog,RS2} have led to the realization that if extra
dimensions exist, they need not be compact. Also ``bulk'' (transverse to
the 3-brane space)  physics turns out to be very interesting giving
alternative possible explanations to other puzzles of particle physics, like the smallness
of the neutrino masses,   the neutrino
oscillations, or the pattern of the SM fermion mass hierarchy (see for 
example \cite{neutr} and references therein).
The phenomenological implication of these constructions are  radical
and the fact that these ideas can, in principle, be put to test in current and
future experiments, makes them very attractive. A comprehensive
account of these ideas can be found in \cite{trieste}.

In  this paper, we present a model which belongs to a  class of brane
universe models  \cite{bigravity,GRS,KR,multi} that suggest that
a part or all of gravitational interactions come from massive
gravitons. In the first case (``bigravity''), gravitational interactions
are the net effect of a massless graviton and a finite number of KK
states that have sufficiently small mass (or/and coupling) so that
there is
no conflict with phenomenology. 
In the second case (``multigravity''), in the absence of a massless
graviton, the normal Newtonian gravity at intermediate scales is reproduced by
special properties of the lowest part of the KK tower (with mass close to zero). These models predict that at
sufficiently large scales, which correspond to the Compton wavelength
of the KK states involved, 
modifications to either  the Newtonian coupling constant or  the inverse
square law will appear due to the Yukawa suppression of the KK states
contribution. These modifications affect the  CMB power spectrum and
thus are, in principle, testable \cite{cmb}. 

The prototype model of this class was the $^{\prime \prime
}+-+^{\prime \prime }$ ``bigravity'' model \cite{bigravity}  where we considered a modification of the 
compact Randall-Sundrum model \cite{RS1} (RS1) with two positive tension flat branes ($''+''$
branes) separated by one intermediate negative tension flat brane ($''-''$
brane) in $AdS_5$ bulk. The task of finding the KK spectrum reduces 
to a simple quantum mechanical problem. It is simple to see that the
model (as every compact model) has a massless graviton that
corresponds to the ground state of the system whose wavefunction
follows the ``warp'' factor. It is also easy to see as in figure 1 that there
should be a state with   wavefunction antisymmetric with respect to
the minimum of the ``warp'' factor, whose mass splitting from the massless
graviton will be very small compared to the ones of the
higher levels. Because the ``warp'' factor is exponential the difference
of the mass behaviour of the first and the rest of the KK states
is also exponential. This allowed for the construction of  a
``bigravity'' model in which the remainder of the KK tower does not
affect gravity beyond the millimeter bound. In
this case gravity at large  scales beyond the Compton wavelength of
the first state will be modified in the sense that
Newton's constant will be reduced. In particular, it is reduced by one
half in the symmetric configuration and it can be effectively switched
off in the highly asymmetric case.

\begin{figure}[t]
\begin{center}
\begin{picture}(300,200)(0,50)

\SetWidth{2}
\Line(10,50)(10,250)
\Line(290,50)(290,250)

\SetWidth{0.5}
\Line(150,50)(150,250)
\Line(10,150)(290,150)

\Text(-10,250)[c]{$''+''$}
\Text(310,250)[c]{$''+''$}
\Text(170,250)[c]{$''-''$}


\Curve{(10,240)(50,192)(65,181)(80,173)(95,167)(110,162)(130,157)(150,155)}
\Curve{(150,155)(170,157)(190,162)(205,167)(220,173)(235,181)(250,192)(290,240)}


\DashCurve{(10,240)(50,192)(65,181)(80,173)(95,167)(110,161)(130,155)(150,150)}{4}
\DashCurve{(150,150)(170,145)(190,139)(205,133)(220,127)(235,119)(250,108)(290,60)}{4}


\DashCurve{(10,145)(15,149)(20,150)(40,153)(65,158)(140,209)(150,210)}{1}
\DashCurve{(150,210)(160,209)(235,158)(260,153)(280,150)(285,149)(290,145)}{1}

\end{picture}
\end{center}

\caption{The graviton (solid line), first (dashed line) and second
(dotted line) KK states wavefunctions in the symmetric $''+-+''$ model. The
wavefunctions are not smooth on the $''-''$ branes. The same pattern
have also the $''++''$ model wavefunctions with the position of the
$''-''$ brane corresponding to the minimum of the warp factor. The
wavefunctions are then smooth.}
\label{wfunct}

\end{figure}
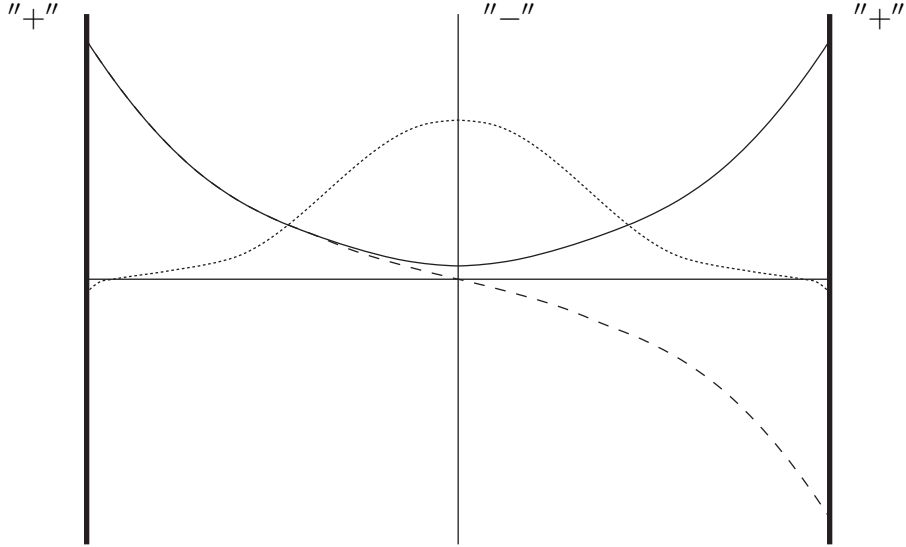

Independently, Gregory, Rubakov and Sibiryakov \cite{GRS} (GRS) suggested  a construction in which
gravity is also modified at ultra-large scales. The GRS
model modifies the decompactified Randall-Sundrum model \cite{RS2} (RS2) by adding a $^{\prime
\prime }-^{\prime \prime }$ brane of half the tension of the $^{\prime
\prime }+^{\prime \prime }$ brane and requiring flat space to the right of
the new $^{\prime \prime }-^{\prime \prime }$ brane. This model does not
have a normalizable 4D graviton but generates 4D gravity at intermediate
distances due to a resonance-like behaviour \cite{reso,dvali1} of the wavefunctions of the KK
states continuum. Gravity in this picture appears to be ``quasi-localized''
and this time it is modified at  ultra-large scales in the sense that
the inverse square law becomes inverse cube, \textit{i.e.} five
dimensional. Although the $^{\prime \prime
}+-+^{\prime \prime }$  and GRS models look quite different it was
shown in \cite{KR} and \cite{multi} that these two models are the limiting cases of a
more general $^{\prime \prime }+--+^{\prime \prime }$ multi-brane model that
interpolates between the ``bigravity'' $^{\prime \prime
}+-+^{\prime \prime }$  model and the
``quasi-localized'' gravity GRS model. The key element of both models
that made ``multigravity'' possible was the bounce of the ``warp'' factor
generated by the $''-''$ branes.

However, one should be careful when dealing with models with
ultralight massive KK states because it is known  that 
 the extra polarizations of the massive gravitons do not
decouple in the limit of vanishing mass, the famous van Dam
- Veltman - Zakharov \cite{VZ}
discontinuity. This could make these models disagree \cite{dvali1}
with standard tests of General Relativity, as for example the bending of
the light by the sun. Furthermore, the moduli (radions) associated with the perturbations of the
$''-''$ branes are necessarily physical ghost fields \cite{ghost}, therefore
unacceptable. The latter problem is connected to the violation of the
weaker energy condition \cite{weak} on $''-''$ branes sandwiched between $''+''$
 branes.  In the GRS model this radion cancels \cite{rest} the extra
polarizations of the massive gravitons and gives the graviton
propagator the correct tensorial structure at intermediate
distances. However, the model has still an explicit ghost in the
spectrum which reveals itself as scalar antigravity at cosmological
scales. A mechanism of
cancelling  \textit{both} the extra massive graviton polarizations \textit{and} the
radion field contribution was suggested in \cite{KR,multi} and
involves some bulk dynamics which are necessary to stabilize
the system, based on a scenario described in \cite{cosmol1}. This mechanism is however non-local in the extra
dimension and because of this may not be very attractive. 

In the present paper we will demonstrate that there is actually a way
out of both these problems. We will use a two brane model with only
$''+''$ branes which was known in \cite{kaloper,nihei,kim,freed,cosmol1,cosmol2,cosmol3} to exhibit a bounce
of the ``warp'' factor, and therefore is bound to have ``bigravity''
by general arguments presented in \cite{bigravity,ktalk}. This can 
be achieved if we consider two $AdS_4$ branes in $AdS_5$
bulk. Motivated by quasi-localization of gravity, Randall and Karch
\cite{KaR} studied the non-compact case of a single $''+''$
brane, which is a limiting case of the asymmetric $''++''$ model. The
weaker energy condition is satisfied so there is 
no ghost modulus in this setup. Furthermore, as was shown in
\cite{limit}\footnote{One day after  this paper had appeared in the hep-archives,
ref.\cite{por} appeared, reaching the same conclusions by a
different method.}, in $AdS$ space it is possible to  circumvent the
van Dam - Veltman - Zakharov no go theorem about the non-decoupling of 
the massive graviton extra polarization states. The price we pay
is that there is a remnant negative cosmological constant on the
brane. This sets an horizon scale and unfortunately the Compton
wavelength of the  light state of  the system lies exponentially far
from this horizon. This does not change even if we consider a highly asymmetric
version of this model. As a result, this ``bigravity'' model makes
no predictions for observable deviations from Newtonian gravity at
ultra-large distances. In addition, although theoretically there exist
modifications of  General Relativity  at all scales due to the
additional polarization states of the massive graviton, they are 
so highly suppressed that they are not observable. Additionally, let
us note that recently models with only positive tension branes and similar ``warp'' factors (created in a different and
dynamical way) were discussed in \cite{nickthegreek,odintsov,selivanov}.

\sect{The two positive brane model}

The model consists of two  3-branes with tensions $V_{1}$ and $V_{2}$ respectively, in an $AdS_5$ space with
five dimensional cosmological constant $\Lambda<0$. The $5$-th dimension has the geometry
of an orbifold and the branes are located at its fixed points,
\textit{i.e.}  $L_0=0$ and $L_{1}=L$. Due to orbifolding, we can
restrict ourselves to the region $0\leq{z}\leq{L}$, imposing the suitable boundary conditions to our solutions. Firstly, we  find a suitable vacuum solution. The action of this setup is:
\be
S=\int d^4 x \int_{-L}^{L} dz \sqrt{-G} 
\{-\Lambda + 2 M^3 R\}+\sum_{i}\int_{z=L_i}d^4xV_i\sqrt{-\hat{G}^{(i)}}
\ee
where $\hat{G}^{(i)}_{\mu\nu}$ is the induced metric on the branes. The notation is the same as in
Ref. \cite{RS1}. The Einstein equations that arise from this
action are:
\be
R_{MN}-\frac{1}{2}G_{MN}R=-\frac{1}{4M^3}
\left(\Lambda G_{MN}-
\sum_{i}V_i\frac{\sqrt{-\hat{G}^{(i)}}}{\sqrt{-G}}
\hat{G}^{(i)}_{\mu\nu}\delta_M^{\mu}\delta_N^{\nu}\delta(z-L_i)\right)
\ee

In order to find a specific form of the equations of motion we need to
write a metric ansatz which will take in account the spacetime symmetries of the
3-brane. Since we would like not to restrict our model to flat solutions on the
branes, we should make a choice which will let us interpolate between
the maximally symmetric space-times in four dimensions, \textit{i.e.} de-Sitter,
Minkowski and Anti-de-Sitter. The metric ansatz \cite{kaloper} that accomplishes this is the following:
\be
ds^2=a^{2}(z)(-dt^{2}+e^{2Ht}d\vec{x}^{2}) +b^{2}(z)dz^2
\ee
where $H$ is the ``Hubble'' parameter and is determined in terms of the brane
tension $V_{1}$ and the bulk cosmological constant  $\Lambda$ from
Einstein's equations. The $z$-dependent function $a(z)$ is the ``warp''
factor that is essential for the localization of gravity and also for
producing the hierarchy between the two branes. 
 In the  case of flat brane solution, \textit{i.e.} the
effective cosmological constant on the brane is zero, we have $H=0$. On
the other hand if we demand a de-Sitter solution on the brane,
\textit{i.e.} the effective cosmological constant on the branes is
positive, we have
$H^{2}>0$. In the case of Anti-de-Sitter solution, \textit{i.e.} the effective
cosmological constant on the branes is negative,  we have $H^{2}<0$
and thus $H$ is imaginary. In order to get a physical 
interpretation of the latter case it is necessary to analytically continue the solution by
a coordinate transformation of the form $t=-ix_1'$, $x_1=it'$, $x_2=x_2'$
and $x_3=x_3'$. After this transformation the metric ansatz can be written
in the  
following form:
\be
ds^2=a^{2}(z)(dx_1'^{2}+e^{2Hx_1'}(-dt'^2+dx_2'^2+dx_3'^2)) +b^{2}(z)dz^2
\ee

Furthermore, in order to have a more compact notation for all cases of maximally
symmetric spaces and
simplify our calculations, it is useful to bring the metric ansatz in the
form:

\be
ds^2=\frac{a^2(z)}{(1-\frac{H^{2}x^{2}}{4})^2}\eta_{\mu\nu}dx^{\mu}dx^{\nu} +b^2(z)dz^2
\ee
where $x^2=\eta_{\mu\nu}x^{\mu}x^{\nu}$. It can be shown that the
Ricci scalar for this metric is $R=-12H^2$. Thus  this metric represents all
maximally symmetric spaces: Minkowski for $H^2=0$, Anti-de-Sitter for
$H^2>0$ and de-Sitter for $H^2<0$. From
now on we shall choose the gauge $b(z)=1$ where our coordinate system
is Gaussian Normal. 
A straightforward
calculation of the Einstein's equations gives us the following differential equations for $a(z)$:
\ba
a'^2(z)&=&H^2-\frac{\Lambda}{24M^3}a^2(z)\\
a ''(z)&=&-\sum_{i}\frac{V_i}{12M^3}a(z)\delta(z-L_i)-\frac{\Lambda}{24M^3}a(z)
\ea
By solving the above equations we find that the solution can be written
in the form :

\be
a(z)=\cosh(k|z|)+\frac{V_{1}k}{\Lambda}\sinh(k|z|)
\ee
with
\be
\renewcommand{\arraystretch}{1.5}
|H^2|=\left\{\begin{array}{cl}\frac{k^2}{\Lambda^2}(V_{1}^{2}k^2-\Lambda^2)&,\frac{|\Lambda|}{k}<V_{1}
~~{\rm for}~dS_{4}~{\rm branes}\\
0&,\frac{|\Lambda|}{k}=V_{1}~~{\rm for~flat~branes}\\
\frac{k^2}{\Lambda^2}(\Lambda^2-V_{1}^{2}k^{2})&,\frac{|\Lambda|}{k}>V_{1}~~{\rm for}~AdS_{4}~{\rm branes}\end{array}\right.
\
\ee
where we have normalized $a(0)=1$ and assumed $V_{1}>0$. Also we have
defined $k\equiv{\sqrt{\frac{-\Lambda}{24 M^3}}}$.

Additionally, in order to have this solution, the brane tensions
$V_{1}$, $V_{2}$, the bulk cosmological
constant $|\Lambda|$ and the position of the second brane $L$ must be
related through the equation: 

\be
\tanh(kL)=k|\Lambda|\frac{V_{1}+V_{2}}{|\Lambda|^{2}+k^{2}V_{1}V_{2}}
\label{ten}
\ee

Let us now restrict ourselves to the case of $AdS_{4}$ spacetime  on the two
branes which will turn out to be the most interesting. In this case
the condition $\frac{|\Lambda|}{k}>V_{1}$ must hold. Hence, we can
define $\tanh(kz_{0})\equiv\frac{k V_{1}}{|\Lambda|}$ and
write the solution in the form: 

\be
a(z)=\frac{\cosh(k(z_{0}-|z|))}{\cosh(kz_{0})}
\ee
from which it is clear that the ``warp'' factor has a minimum at
$z=z_{0}$. From this point we can see the role of the
$AdS_{4}$ on the branes, \textit{i.e.} the role of the condition
$\frac{|\Lambda|}{k}>V_{1}$. This condition allows us to have the bounce
form of the ``warp'' factor (\textit{i.e.} a minimum in the ``warp''
factor) allowing the second brane to have
positive tension and give us, as we will see shortly, a
 phenomenology quite similar to the $^{\prime \prime
}+-+^{\prime \prime }$ ``bigravity'' model \cite{bigravity}.
This can be easily seen from the eq.(\ref{ten}) which relates the brane
tensions and the distance between the branes. From this equation we indeed
see that by placing the second brane
after the minimum of the ``warp'' factor we can make the tension of
the second brane positive and thus both branes that appear in the model
have positive tension avoiding the problems associated with 
negative tension branes. 
In fact it is clear that the present model mimics the characteristics
of the $^{\prime \prime
}+-+^{\prime \prime }$ ``bigravity'' model since what we effectively do is to reproduce the
effect of the presence of a negative tension brane, \textit{i.e.} the bounce form
of the ``warp'' factor, with another mechanism allowing a negative four
dimensional cosmological constant on the brane. Note that in the
limit that ${V_{1}}\rightarrow\frac{|\Lambda|}{k}$ (flat limit) the
minimum of the ``warp'' factor tends to infinity and if we wish to
have a brane at a finite point, it will necessarily have negative
tension.

The relationship between the 4D effective fundamental scale  $M_{*}$\footnote{the factor $2M^{2}_{*}$ multiplies the four dimensional Ricci scalar
in the Lagrangian after dimensionally reducing} and the five dimensional
fundamental scale $M$ can be easily found by dimensional reduction to be:

\be
M_{\rm
*}^2=\frac{M^3}{k\cosh^{2}(kz_{0})}\left[kL+\sinh(kL)\cosh(k(L-2z_{0}))\right]
\label{plank}
\ee

The above formula tells us that for finite $L$ the compactification
volume is finite and thus the zero mode is normalizable. In the case
where we send the second brane to infinity, the compactification
volume becomes infinite which indicates that the zero mode becomes
non-normalizable. Note that $M_{*}$ is not
necessarily equal to $M_{Pl}$ since as will see shortly, at least for
a sector of the parameter space of our model, gravity is the result not
only of the massless graviton but also of an ultralight KK state.

The ``warp'' factor renormalizes the physical scales of the theory as
 in \cite{RS1}. Thus, all 
mass parameters $m_0$ on a brane placed at the point $z$ are rescaled as

\be
m=a(z)m_{0}
\ee

Hence, assuming some kind of stabilization mechanism which fixes the
positions of the branes, one can choose  a distance between the
two branes such that this rescaling  leads to the creation of a desired mass
hierarchy.

However, since we consider non-flat solutions on the branes, we
have to make sure that the four dimensional effective cosmological
constant does not contradict  present experimental and
observational bounds. Recent  experimental data favour a positive
cosmological constant, nevertheless since
zero cosmological constant is not completely ruled out it can be argued that also
a tiny negative cosmological constant can be acceptable within the experimental
uncertainties. The effective cosmological constant on the two branes
is:

\be
\Lambda_{4d}=-12H^{2}M_{*}^2=-\frac{12}{\cosh^2(kz_{0})}k^{2}M_{*}^{2}
\ee
   
From the above formula we can see that we can make the cosmological
constant small enough $|\Lambda_{4d}| \la 10^{-120} M_{\rm Pl}^4$ if we
choose large enough $kz_{0}$, \textit{i.e.} $kz_{0}\ga {135}$. This
however will make observable deviations from Newtonian gravity at
ultra-large scales impossible as we will see in the 
next section.

To determine the phenomenology of the model we need to know the KK
spectrum that follows from the dimensional reduction. This is
determined by considering the (linear) fluctuations of the metric
around the vacuum solution that we found above. We can write the
metric perturbation in the form:

\be
ds^2=\left[a(z)^{2}g^{AdS}_{~\mu\nu} +\frac{2}{M^{3/2}}h_{\mu\nu}
(x,z)\right]dx^\mu dx^\nu +dz^2
\ee
where $g^{AdS}_{~\mu\nu}$ is the vacuum solution. Here we have ignored the radion mode that could be used to stabilize
the brane positions $z=L_{0}$ and $z=L_{1}$, assuming some
stabilization mechanism. We
expand the field $h_{\mu\nu}(x,z)$ into graviton and KK plane waves:
\be
h_{\mu\nu}(x,z)=\sum_{n=0}^{\infty}h_{\mu\nu}^{(n)}(x)\Psi^{(n)}(z)
\ee
where we demand
$\left(\nabla_\kappa\nabla^\kappa +2 H^2-m_n^2\right)h_{\mu\nu}^{(n)}=0$
and additionally 
\mbox{$\nabla^{\alpha}h_{\alpha\beta}^{(n)}=h_{\phantom{-}\alpha}^{(n)\alpha}=0$}.
The function $\Psi^{(n)}(z)$ will obey a second order differential
equation which after a change of variables and a redefinition of the wavefunction reduces to an ordinary
Schr\"{o}dinger-type equation:
\be
\left\{-
\frac{1}{2}\partial_w^2+V(w)\right\}\hat{\Psi}^{(n)}(w)=\frac{m_n^2}{2}\hat{\Psi
}^{(n)}(w)
\ee
where the potential is given by:

\ba
V(w)=&-&
\frac{9\tilde{k}^{2}}{8}~+~\frac{15\tilde{k}^2}{8}\frac{1}{\cos^{2}\left(\tilde{k}(|w|-w_{0})\right)}\cr      &-&\frac{3k}{2}\left[ \tanh(kz_{0})\delta(w)+\frac{\sinh(k(L-z_{0}))}{\cosh(kz_{0})}
\delta(w-w_{1})\right] 
\ea
with $\tilde{k}$  defined as
$\tilde{k}\equiv{\frac{k}{\cosh(kz_{0})}}$. The new variables  and the redefinition of the wavefunction are
related with the old ones by:
\be
w\equiv {\rm sgn}(z)\frac{2}{\tilde{k}}\left[\arctan\left(\tanh(\frac{k(|z|-z_{0})}{2})\right)+\arctan\left(\tanh(\frac{kz_{0}}{2})\right)\right]
\
\ee
\be
\hat{\Psi}^{(n)}(w)\equiv\frac{1}{\sqrt{a(z)}}\Psi^{(n)}(z)
\ee

Thus in terms of the new coordinates, the branes are  at $w_{L_{0}}=0$
 and $w_{L}$, with the minimum of the potential  at $w_{0}={2 \over \tilde{k}}\arctan\left(\tanh(\frac{kz_{0}}{2})\right)$. Also note
that with this transformation the point $z=\infty$ is mapped to the
finite point $w_{\infty}={2 \over \tilde{k}}\left[{\pi \over 4} + \arctan\left(\tanh(\frac{kz_{0}}{2})\right)\right]$.

From now on we restrict ourselves to the symmetric configuration of the two
branes with respect to the minimum  $w_{0}$ (\textit{i.e.} the first
brane at 0 and the
second at  $2w_{0}$ ), since the important characteristics of the model
appear independently of the details of the configuration.
Thus, the model has been reduced to a ``quantum mechanical problem''
with $\delta$-function potentials wells of
the same weight and an extra smoothing term in-between (due to the AdS
geometry). This  gives the potential a double ``volcano'' form. 

An interesting characteristic of this potential is that it always (for
the compact cases \textit{i.e.} $w_{L}<w_{\infty}$)
gives rise to a normalizable massless zero mode, something that is
expected since the volume of the extra dimension is finite.  The zero
mode wavefunction is given by:
\be
\hat{\Psi}^{(0)}(w)=\frac{A}{[\cos(\tilde{k}(w_{0}-|w|))]^{3/2}}
\ee
where the normalization factor $A$ is determined by the requirement 
$\displaystyle{\int_{-w_{L}}^{w_{L}}
dw\left[\hat{\Psi}^{(0)}(w)\right]^2=1}$, chosen so that we get the standard 
form 
of the Fierz-Pauli Lagrangian.

The form of the zero mode resembles the one  of the zero mode of the $''+-+''$
model, \textit{i.e.} it has a bounce form with the turning at $w_{0}$ 
(see figure 1). In the
case of the $''+-+''$ the cause for this was the presence of the
$''-''$ brane. In the present model it turns out that by considering
$AdS$ spacetime on the branes we get the same effect.

In the case that we send the second brane to infinity
(\textit{i.e.} $w\rightarrow {w_{\infty}}$) the zero mode fails to be normalizable
due to singularity of the wavefunction exactly at that point. This can
be also seen from eq.(\ref{plank}) which implies that at this limit $M_{*}$
becomes infinite (\textit{i.e.} the coupling of the zero mode becomes zero). Thus in this limit the
model has no zero mode and all gravitational interactions must be
produced by the ultralight first KK mode. The spectrum in this
case was discussed by Randall and Karch at \cite{KaR}. 

Considering the Schr\"{o}dinger equation for $m\ne0$ we can determine
the wavefunctions of the KK tower. It turns out that the differential
equation can be brought to a hypergeometric form, and hence  the
general solution  is given in terms two hypergeometric functions: 

\be
\renewcommand{\arraystretch}{1.5}
\begin{array}{c}\hat{\Psi}^{(n)}=\cos^{5/2}(\tilde{k}(|w|-w_{0}))\left[C_{1}~F(\tilde{a}_{n},\tilde{b}_{n},\frac{1}{2};\sin^{2}(\tilde{k}(|w|-w_{0})))~~~~~~~~\right.
\\ \left.  ~~~~~~~~~~~~~~~~~~~+C_{2}~|\sin(\tilde{k}(|w|-w_{0}))|~F(\tilde{a}_{n}+\frac{1}{2},\tilde{b}_{n}+\frac{1}{2},\frac{3}{2};\sin^{2}(\tilde{k}(|w|-w_{0})))\right]
\end{array}
\ee
where
\ba
\tilde{a}_{n}=\frac{5}{4}+\frac{1}{2}\sqrt{\left(\frac{m_{n}}{\tilde{k}}\right)^2+\frac{9}{4}}
\cr \tilde{b}_{n}=\frac{5}{4}-\frac{1}{2}\sqrt{\left(\frac{m_{n}}{\tilde{k}}\right)^2+\frac{9}{4}}
\ea
The boundary conditions (\textit{i.e.} the jump of the wave function at the points
$w=0$, $w_{L}$) result in a
$2\times2$ homogeneous linear system which, in order to have a
non-trivial solution, leads to the vanishing determinant. In the
symmetric configuration which we consider, this procedure can be
simplified by considering even and odd functions with respect to the
minimum of the potential $w_{0}$.

In more detail, the odd eigenfunctions obeying
the  b.c. $\hat{\Psi}^{(n)}(w_{0})=0$  will  have $C_1=0$ and thus the form:
\be
\hat{\Psi}^{(n)}=C_{2}\cos^{5/2}(\tilde{k}(|w|-w_{0}))|\sin(\tilde{k}(|w|-w_{0}))|~F(\tilde{a}_{n}+\frac{1}{2},\tilde{b}_{n}+\frac{1}{2},\frac{3}{2};\sin^{2}(\tilde{k}(|w|-w_{0})))
\ee

On the other hand, the even eigenfunctions obeying
the b.c. $\hat{\Psi}^{(n)}~'(w_{0})=0$  will have $C_2=0$ and thus the form:

\be
\hat{\Psi}^{(n)}=C_{1}\cos^{5/2}(\tilde{k}(|w|-w_{0}))F(\tilde{a}_{n},\tilde{b}_{n},\frac{1}{2};\sin^{2}(\tilde{k}(|w|-w_{0})))
\ee

The remaining boundary condition is given by:
\be
\hat{\Psi}^{(n)}~'(0)+\frac{3k}{2}\tanh(kz_{0})\hat{\Psi}^{(n)}(0)=0
\ee
and determines the mass spectrum of the KK states. From this
quantization condition we get that the KK spectrum has a special first
mode similar to the one of the $^{\prime \prime
}+-+^{\prime \prime }$ ``bigravity'' model. For $kz_0 \ga 5$  the  mass 
of the first mode is given by the approximate relation:
\be
m_1=4\sqrt{3}~k~e^{-2kz_{0}}
\label{m1}
\ee

In contrast, the masses of the next levels, if we  put together the
results for even and odd wavefunctions, are given by the formula:
\be
m_{n+1}=2\sqrt{n(n+3)}~k~e^{-kz_{0}}
\label{mr}
\ee
with $n=1,2,...$.

We note that the first KK state has a different scaling law with respect
to the position of the minimum of the ``warp'' factor compared
to the rest of the KK tower, since it scales as $ e^{-2kz_{0}}$ while
the rest of the tower scales as $e^{-kz_{0}}$. Thus the first
KK state is generally much lighter than the rest of the tower. It is clear that
this mass spectrum resembles the one of the $^{\prime \prime
}+-+^{\prime \prime }$ ``bigravity'' model. The deeper
reason for this is again the common form of the ``warp'' factor. In both
cases the ``warp'' factor has a minimum due to its ``bounce''
form. The graviton wave function  follows
the form of the ``warp'' factor, \textit{i.e.} it is symmetric with respect to $w_{0}$, while
 the wavefunction of the first KK state is antisymmetric in
respect to $w_{0}$ (see figure 1). The absolute values of the two wavefunctions  are almost identical in all
regions except near $w_{0}$ (the symmetric is nonzero while the
antisymmetric is zero at $w_{0}$). The graviton
wavefunction is suppressed by the factor
$\frac{1}{cosh^{2}(kz_{0})}$ at $w_{0}$
which brings it's value close to zero for reasonable values of
$kz_{0}$. Thus, the mass
difference which is determined by the wavefunction near  $w_{0}$ is expected  to be generally very small, a fact which formally appears as
the extra suppression factor $e^{-kz_{0}}$ in the formula of $m_{1}$
in comparison with the rest of the KK tower. 

In the case that we consider an asymmetric brane configuration,
for example
$w_{L}>2w_{0}$ the spectrum is effectively independent of the position
of the second brane $w_{L}$ (considering $kz_{0}\ga 5$). Thus, even in the case that we place the
second brane at $w_{\infty}$, \textit{i.e.} the point which corresponds to infinity in the
$z$-coordinates, the spectrum is given essentially by the same
formulas. In the case that the second brane is placed at
$w_{0}<w_{L}<2w_{0}$,  some dependence on the position of the second
brane (\textit{i.e.} dependence on the scale hierarchy between the branes) is
present. Nevertheless, the main characteristics of the spectrum remain
the same, \textit{i.e.} the first KK state is special and always much lighter
than the others. In conclusion, the key parameter which determines the
spectrum is the position of the minimum of the ``warp'' factor.

Returning to our wavefunction solutions, we should note that for each eigenfunction the normalization constants $C_{1}$ and $C_{2}$
can be determined by the normalization condition 
$\displaystyle{\int_{-w_{L}}^{w_{L}}dw\left[\hat{\Psi}^{(n)}(w)\right]^2=1}$
 which is such that we get the standard form of the Fierz-Pauli
Lagrangian for the KK states. 
Knowing the normalization of the wavefunctions, it is straightforward
to calculate
the strength of the interaction of the KK states with the SM
fields confined on the brane\footnote{In the symmetric configuration it does not make any
difference which brane is our universe.}. This can be calculated  by expanding the minimal
gravitational coupling of the SM Lagrangian $\displaystyle{\int
d^4x\sqrt{-\hat{G}}{\mathcal{L}}\left(\hat{G},SM fields\right)}$ with respect to 
the metric. In this way we get:

\ba
{\mathcal{L}}_{int}&=&-\frac{1}{M^{3/2}}\sum_{n\geq
0}
\hat{\Psi}^{(n)}\left(w_{\rm brane}\right)h_{\mu\nu}^{(n)}(x)T_{\mu\nu}\left(x\right)= 
\nonumber
\\&=&-\frac{A}{M^{3/2}}h_{\mu\nu}^{(0)}(x)T_{\mu\nu}\left(x\right)-
\sum_{n>0}\frac{\hat{\Psi}^{(n)}\left(w_{\rm brane}\right)}{M^{3/2}
}h_{\mu\nu}^{(n)}(x)T_{\mu\nu}\left(x\right)
\ea
with $T_{\mu\nu}$ the energy momentum tensor of the SM
Lagrangian. Thus, the coupling of the zero and the first KK
mode to matter are respectively:
\ba
\frac{1}{c_0}&=&\frac{A}{M^{3/2}}=\frac{1}{M_{*}} \label{c1} \\
\frac{1}{c_1}&=&\frac{\hat{\Psi}^{(1)}\left(w_{\rm brane}\right)}{M
^{3/2}}\simeq\frac{1}{M_{*}} \label{c2}
\ea
where $A$ is the zero mode normalization constant which turns out
to be $\frac{M^{3/2}}{M_{\rm *}}$. We should also note that the
couplings of the rest of the KK states are much smaller and scale as $e^{-kz_{0}}$.

Exploiting the different mass scaling of the first KK relative to the
rest we can ask whether it is possible to realize a ``bigravity''
scenario similar to that in $^{\prime \prime
}+-+^{\prime \prime }$ ``bigravity'' model. 
 In that model by appropriately choosing the position of the minimum of the ``warp''
factor, it was possible to make the first KK state have mass such that the
corresponding wavelength is  of the order of the cosmological scale that gravity
has been tested and at the same time have the rest of the KK tower wavelengths
below 1mm  (so that there is no conflict with Cavendish bounds). In
this scenario the gravitational interactions are due to the net effect of the massless graviton and the first ultralight KK
state. From eq.(\ref{c1}), (\ref{c2}) it can be understood that in the symmetric
configuration the massless graviton and the special KK state
contribute by the same amount to the gravitational interactions.
In other words:
\be
\frac{1}{M_{Pl}^2}=\frac{1}{M_{*}^2}+\frac{1}{M_{*}^2}~~~ \Rightarrow~~~ M_{Pl}=\frac{M_{*}}{\sqrt{2}}
\ee

 In the present model, the fact that the effective four dimensional
cosmological constant should be set very close to zero, requires that
the  ``warp'' factor is constrained by  \mbox{$kz_{0}\geq{135}$} and thus,
in this case, the spectrum of the KK states will be very dense (tending
to continuum) bringing more states close to zero mass. The KK states
that have masses which correspond to wavelengths larger than $1mm$ have
sufficiently small coupling so that there is no conflict with 
phenomenology (the situation is exactly similar to the RS2 case
where the coupling of the  KK states is proportional to their mass and thus
it is decreasing for lighter KK states).  The fact that the spectrum
tends to a continuum shadows
the special role of the first KK state. Moreover, it is interesting to note that
at the limit where the minimum of the ``warp'' factor is sent to infinity ($w_{\infty}$)
 the special behaviour of the first KK persists and does not catch the
other levels (by changing its scaling law) as
was the case in \cite{multi}. This means that the limit  $w\rightarrow
w_{\infty}$ will indeed
be identical to \textit{two}  RS2, but on the other hand it is interesting to note
that what we call graviton in the RS2 limit is actually the
combination of a massless graviton
\textit{and} the ``massless'' limit of the special first KK
state. This ``massless'' limit exists as we will see in the next
section and  ensures that locality is respected by the
model, since physics on the brane does not get affected from the
boundary condition at infinity.

\section{Discussion and conclusions}

The fact that we have a ultralight graviton in our spectrum is at
first sight worrying because it is well known that in the flat space
the tensor structure of the propagators of the massless and of the massive
graviton are different \cite{VZ} and that there is no smooth limit between them
when $m \rightarrow 0$. The bending of the light by the sun agrees with 
the prediction of the Einstein theory to $1\%$  accuracy. This
is sufficient to rule out any scenario which a significant component
of gravity is due to a  massive graviton, however
light its mass could be. However, as was shown in \cite{limit}, the situation in $AdS$ space is
quite different. There it was shown
that if we could arrange  ${m_1 \over H} \la 0.1$ there is no
discrepancy with standard tests of Einsteinian gravity as for example
the bending of the light by the sun.

In the particular model we have at hand, it is ${m_1 \over H} \sim
e^{-k z_0}$ so we can easily accommodate the above bound. Then, the
Euclidean propagator (in configuration space) of the massive KK states   for relatively large $z_0$ will be given by:
\be
G^{m}_{\mu\nu;\mu'\nu'}(x,y)=\frac{1}{4 \pi^2
\mu^2}(\delta_{\mu\mu'}\delta_{\nu\nu'}+\delta_{\mu\nu'}\delta_{\nu\mu'}-\left(
1-{1\over 6}e^{-2 kz_0}\right)\delta_{\mu\nu}\delta_{\mu'\nu'})
\ee
where $\mu$ is the geodesic distance between two points. In the above, 
we have omitted terms that do not contribute when integrated with 
a conserved $T_{\mu \nu}$. For $kz_0 \ga 2.3$ there in no problem with the bending of
light. However, if our aim is to see modifications of gravity at
ultra-large distances, this is impossible because the Compton wavelength of our ultralight
graviton will be $e^{kz_0}$ times bigger than the horizon $H^{-1}$ of the
$AdS_4$ space on our brane due to equations (\ref{m1}), (\ref{mr}). The ``Hubble''
parameter follows $m_2$ rather than $m_1$.
 
What happens if one takes  an asymmetric version of this model 
where $L>2z_0$ is that the spectrum  does not get significantly modified so we
are effectively in the same situation. In the 
case where $z_0<L<2z_0$ the spectrum will behave like 
the one of the $''+-+''$ model. Then for $\omega \ll 1$ we will have $m_{1} \sim \omega e^{-2kx} M
\sim e^{-2kx} M_{\rm Pl}$, $H \sim e^{-kz_{0}} M \sim e^{-kx} M_{\rm
Pl}$ where $M \approx  M_{\rm Pl}/\omega$ is the fundamental scale, $\omega \approx e^{-(2z_0-L)}$ is
the ``warp'' factor and
$x=L-z_0$. Again, the  ultralight graviton is hiding well beyond the
$AdS$ horizon. However, the coupling of the remaining of the
KK tower to matter will be different than the symmetric case and one may have different  corrections to  Newton's law  on the left and right branes.     

In summary, in this paper we presented a new ``bigravity'' model with two
$AdS_4$ branes in $AdS_5$ bulk which has a lot of similarities with
the $''+-+''$ ``bigravity'' model. The fact that we have no $''-''$
branes removes the ghost state problem and furthermore, due to some
amazing property of the $AdS$ space we are able to circumvent the van
Dam - Veltman - Zakharov discontinuity of the graviton
propagator. This  makes the model compatible with the predictions of
General Relativity in the small graviton mass limit since  the extra degrees of freedom of the massive
graviton practically decouple. However, the presence of the AdS horizon prevents the
modifications of gravity at large distances becoming
observable. In the future it would be interesting to explore
similar models where it would be possible to obtain observable
modifications of gravity at large (cosmological) scales.  
In particular, it would be interesting to see if the above 
characteristics of this model persist when we add matter density on the
branes. Then it is interesting to examine if/how the ultralight graviton is
going to reveal itself in the cosmological solutions discussed by \cite{cosmol3}.

\textbf{Acknowledgments:} We are
indebted to Graham G. Ross for very important and stimulating
discussions and for careful reading of the manuscript. We would like to thank 
 Joe Bhaseen, Tibault Damour, Nemanja Kaloper, Panagiota Kanti, Keith
Olive, Maxim Pospelov and  Mario G. Santos for useful discussions. We
are grateful to the organizers of the conference  ``Physics beyond four
dimensions'', July 2000, ICTP, Trieste, where ``multigravity'' and the
$''++''$ system were reported \cite{ktalk}, and  Nemanja
Kaloper for important discussions at the early stages of this project.
S.M.'s work is supported by the Hellenic State Scholarship Foundation (IKY) \mbox{No. 
8117781027}. A.P.'s work is supported by the Hellenic State Scholarship
Foundation (IKY) \mbox{No. 8017711802}. This work  is
supported in part by PPARC rolling grant PPA/G/O/1998/00567, by
the EC TMR grants  HRRN-CT-2000-00148 and  HPRN-CT-2000-00152.

\textbf{Addendum:} One day after this work had appeared in the hep-archives,
ref. \cite{KRpaper} appeared, being the published version of \cite{KaR}. In that paper the non-compact case of a single
$''+''$ brane was studied.

\end{document}